\documentclass[referee]{raa}            

\usepackage{graphicx,times}             
\usepackage{natbib}
\usepackage{amssymb,amsmath}
\bibpunct{(}{)}{;}{a}{}{,}

\usepackage{longtable}

\usepackage[pagebackref=true]{hyperref}

\begin{document}

\title{An investigation on the FWHM of absorption features of type Ia supernovae}

\volnopage{Vol.0 (202x) No.0, 000--000}      
\setcounter{page}{1}          

\author{Xulin Zhao  
	\inst{1,*}\footnotetext{$*$  {\it zhaoxulin@139.com},  \href{https://orcid.org/0000-0002-3204-2358}{https://orcid.org/0000-0002-3204-2358}}
	\and Keiichi Maeda
	\inst{2}
	\and Xiaofeng Wang
	\inst{3,*}\footnotetext{$*$ wang\_xf@mail.tsinghua.edu.cn}
}

   \institute{Tianjin Key Laboratory of Quantum Optics and Intelligent Photonics, School of Science, Tianjin University of Technology, Tianjin 300384, China\\
	\and
	Department of Astronomy, Kyoto University, Kyoto, 606-8502, Japan\\
	\and
	Department of Physics, Tsinghua University, Beijing, 100084, China\\
	\vs\no
	{\small Received 202x month day; accepted 202x month day}}

\abstract{We present an investigation of the full width at half maximum (FWHM, or $\gamma$) of absorption features of Type Ia supernova (SNe Ia). We found that, the average value of FWHM can be well predicted with the rest wavelength ($\lambda$). The velocity also plays an important role, as objects with a higher velocity tend to have a larger FWHM. Temperature may be the third factor, as we found that, at the same velocity (but different phases), a normal-velocity (NV) object tends to have a larger FWHM than high-velocity (HV) object. Also, 1991T/1999aa-like objects that are believed to have relatively high temperatures show the largest FWHMs if compared at the same velocity. Generally speaking, FWHM evolves very slowly with time and shows no correlation with $\Delta m_{15}$, but 1991T/1999aa-like objects are characterized by relatively fast decreasing FWHM. On the other hand, we found that, objects with relatively small FWHMs shows a tighter correlation between absorption depth ($A$) and $\Delta m_{15}$, possibly a sign of higher degree of homogeneity. We also found that $A/\gamma$ of Si II $\lambda$5972 has a strong correlation with $\Delta m_{15}$, and more importantly, a relatively slow time evolution, making it a useful luminosity estimator even in the absence of phase information.}


\authorrunning{X. Zhao, K. Maeda \& X. Wang }    
\titlerunning{FWHM of SNe Ia}  

	\maketitle

\section{Introduction\label{S1}}

Type Ia supernovae (SNe Ia) serve as a critical measurement tool in cosmology, with extremely bright and well standardable luminosity \citep[e.g.][]{Phillips93,Phillips99, Riess98, Riess19, Perlmutter99, Freedman19,Jha19}. Physically, SNe Ia are caused by rapid thermonuclear runaway explosion. This usually happens when the white dwarf's mass approached the Chandrasekhar limit ($\approx 1.4 M_{\odot}$), and then electron's degeneracy pressure was not able to counteract the gravitational force \citep[e.g.][]{Nomoto82,Khokhlov91,Hillebrandt00,Maeda10,Maeda18, Flors20}. However, the conditions of explosion are not absolutely strict. For example, sub-Chandrasekhar mass explosion could be possible \citep[e.g.][]{Kushnir13, Pakmor13,Blondin18, Shen18, Flors20}. Progenitor system could be different too: (1) Single degenerate scenario (SD): The progenitor could be a CO white dwarf which accreted mass from a non-degenerate companion star \citep[such as a main-sequence or red-giant star, e.g.][]{Whelan73, Nomoto82,Ruiz-Lapuente23}. (2) Double degenerate scenario (DD): Namely composing two white dwarfs \citep[e.g.][]{Iben84, Webbink84}. A possible event is the famous nearby supernova SN 2011fe. Observations suggested no companion star or dimer than a few percent of the sun \citep[e.g.][]{Li11}.

Due to uncertainties in initial condition, such as those we mentioned above, SN Ia actually has significant intrinsic diversities, and it restricts the accuracy of standardized luminosity. To reduce the deviations, people are looking for spectral indicators that are sensitive to the intrinsic diversity. For example, the velocity of Si II $\lambda$6355 was found to be very helpful in extinction correction for the peak luminosity \citep{Wang09}. Currently the tightest correlation between spectral and photometric indicators is found to be that of absorption depth (i.e. maximum absorption rate) of Si II $\lambda$5972 ($A^{Si5}$) and decline rate $\Delta m_{15}$ \citep[see for example,][]{Zhao21}. But it is still unclear whether this correlation may help improve the calibration of luminosity. An easier method is to restrict the sample to subclass. A convenient classification scheme was proposed by \citet{Wang09}. In this scheme, most objects are included in a `high velocity' (HV) subclass which has a velocity of Si II $\lambda$6355 ($V^{Si6}$) greater than $12,000$ km/s at maximum light, and a `normal velocity' (NV) subclass which has $V^{Si6}\leq 12,000$ km/s at maximum light. Other subclasses include 1991T-like \citep{Filippenko92b,Phillips92}, 1999aa-like \citep{Li01}, 1991bg-like \citep{Filippenko92a}. Type 1991bg-like SNe Ia, on the contrary, undergo a less complete burning (a proof is their low luminosity), leaving more IMEs unburnt in the ejecta, and thus have stronger absorption lines. Type 1991T-like SNe Ia are characterized by weak lines of intermediate-mass elements, as a result of more complete burning. Type 1999aa-like objects have high luminosity but also relatively normal Si II $\lambda$6355 feature. In practice, 1991T-like and 1999aa-like objects are often grouped together as 1991T/1999aa-like objects. The reason is that they are overall similar (main difference is in line strength) with bright luminosity, and also because their samples are often too small \citep[see, e.g.][]{Zhao21}. In Benetti's scheme \citep{Benetti05}, SNe Ia are divided into ‘Faint’, ‘high-velocity gradient (HVG)’, and ‘low-velocity gradient
(LVG)’ subtypes. In Branch's scheme \citep{Branch06, Branch09}, SNe Ia are divided into ‘core-normal (CN)’, ‘broad-lined (BL)’, ‘cool’, and ‘shallow silicon (SS)’.

The absorption strength of an absorption feature is determined by the absorption depth and the full width at half maximum (FWHM). The former represents the absorption rate at the central velocity, while the latter describes how fast the absorption rate decreases at the broadened velocities. The sum of absorption rates gives the `pseudo-equivalent width' ($pEW$) which quantifies the total absorption strength. Compared with the absorption depth and the pseudo-equivalent width, the full width at half maximum is relatively unexplored. But, a better determination of this parameter can significantly improve spectral fittings. Also, its distributions could be useful in researches on the intrinsic diversity, or help constrain the parameters of the explosion models of SNe Ia. Also note that, the absorption feature may include an extra component called `high-velocity feature' \citep[HVF, e.g.][]{Childress14,Maguire14,Silverman15,Zhao15,Meng19,Ni23,Hakobyan25} component. These HVFs are believed to be formed in outer layers, where the ejecta's velocities are higher than the photosphere. Their origins are still a mystery. But, evidences have been found, suggesting possible element transitions \citep[e.g. oxgen to sillicon, see for example][]{Zhao16} and matter transfer from inner shells to outer shells \citep[see for example][]{Zhao24}.

\section{Data and Measurement\label{S2}}

This work targets at regulations of the FWHM of absorption features of SNe Ia. Since the study is relatively general, no special restrictions were imposed on the sample. The spectra were primarily from the CfA supernova program \citep{Matheson08, Blondin12}, the Berkeley supernova program \citep{Silverman12a,Silverman12b}, Carnegie supernova project \citep[CSP,][]{Folatelli13}, and our own database. Also noted that, in this work, the measurements involved only photospheric components, no HVF component.

FWHM is defined as the width of a curve at half of the maximum amplitude. However, in this work, this parameter is not measured directly from the absorption profile. Rather, we obtain it from Gaussian fittings. A Gaussian function is described by $F=Aexp(\frac{-(\lambda-\lambda_0)^2}{2\sigma^2})$, with three fitting parameters. $\lambda$ is the Doppler-shifted wavelength, depending on the velocity and the rest wavelength. `A' describes the absorption depth. Dispersion $\sigma$ is essentially equivalent with FWHM, except for a constant coefficient: $FWHM=2\sqrt{2ln2}\sigma \approx 2.3548\sigma$. It is worth reminding again that, for convenience, we may also use the symbol `$\gamma$' to represent the FWHM, especially in formulas.

Uncertainty from the flux's noise ($U_{flux}$) is usually about 1\% $\sim$ 2\%, i.e. a 50 $\sim$ 100 Signal/Noise ratio. This uncertainty does not completely transmit to the final result, because the noise was partly smoothed off. For velocity and $\gamma$, the efficiency should be even lower, because they are not directly affected by the flux. Rather, they are determined by the wavelength which is in the perpendicular direction. For these reasons, the efficiency ($r_{flux}$) could be $\approx0\%$ for $V$ and $\gamma$. But to be strict, in this work it is set to be 30\%. For $A$ which is directly affected by the flux, we set $r_{flux}=100\%$. The other part of uncertainty ($U_{fit}$) is given by fitting program in Matlab program. Total uncertainty is calculated by $U=\sqrt{(r_{flux}U_{flux})^2+U_{fit}^2}$. 

Fitting for velocity is better guided by the maximum absorption, so its uncertainty is relatively low at $<1$\% level, unless the pseudo-continuum was poorly determined. Fitting uncertainty of the absorption depth is usually at a level $<2$\%. Uncertainty of FWHM is relatively higher, partly because the line blending or non-Gaussian profile. Usually it is at a level $<5$\%. Anyway, our main targets are the trends or possible correlations, not on accurate calculation or individual object, so the errors are not really important in this study.



\section{Time evolution of FWHM\label{S3}}


Physically, absorption depths are mainly decided by luminosity, the ion abundance, and the energy levels. In comparison, FWHM is majorly affected by velocity vectors (note that direction also matters) in the ejecta, which decides the Doppler broadening. The velocity vectors were at least affected by temperature and collisions. These factors were likely almost determined in the explosion, barely changed afterward.

\begin{figure*}
	\centering
	\includegraphics[width=1\columnwidth]{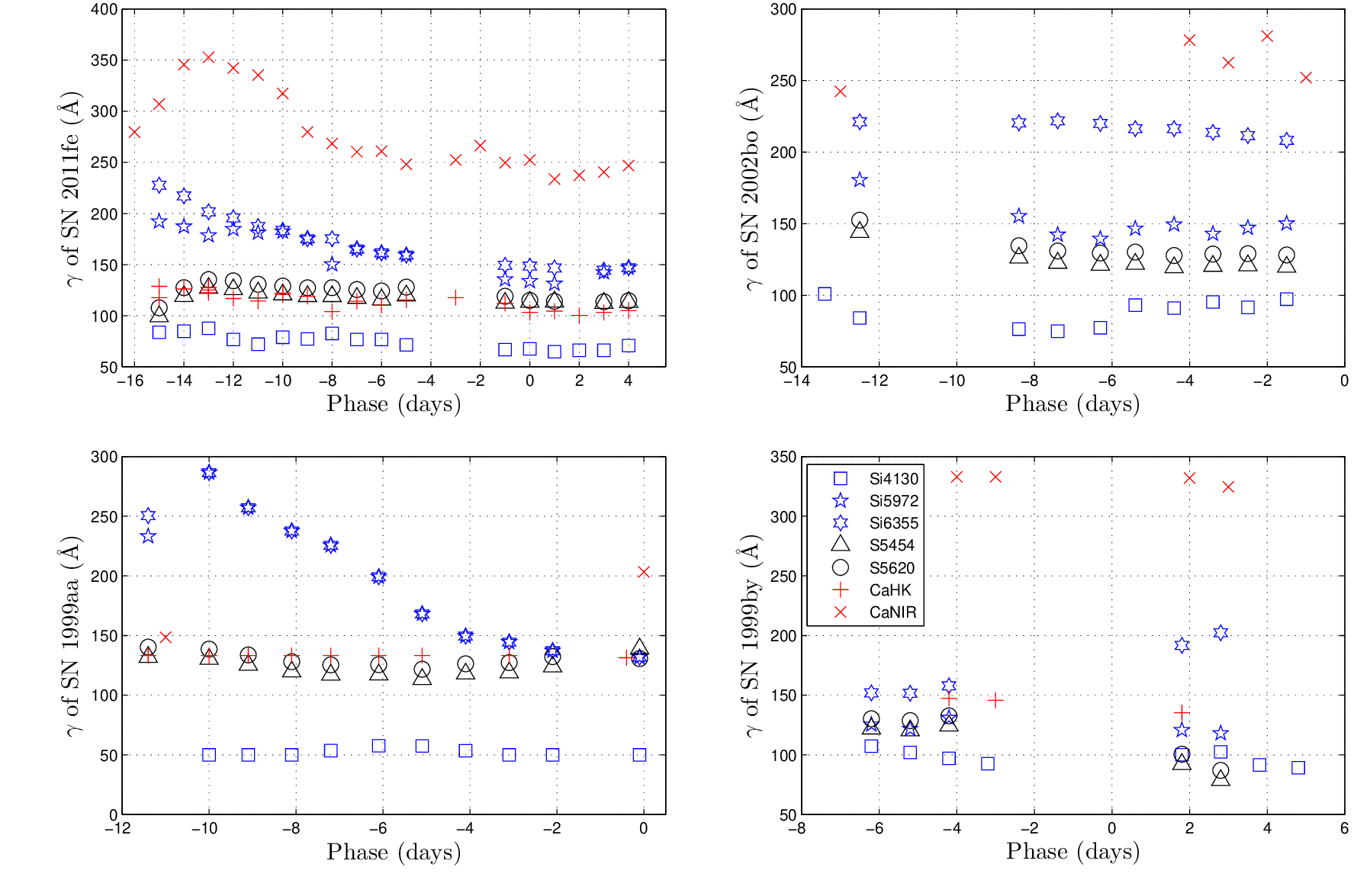}
	\caption{\label{Fig1} Time evolutions of the FWHMs ($\gamma$) of SN 2011fe (upper left panel, NV subtype), SN 2002bo (upper right panel, HV subtype), SN 1999aa (lower left panel, 1991T/1999aa subtype) and SN 1999by (lower right panel, NV subtype). Marker specifiers: the square, pentagram, hexagram, triangle, plus and cross mark the absorption features Si II $\lambda$4130, Si II $\lambda$5972, Si II $\lambda$6355, S II $\lambda$5454, S II $\lambda$5620, Ca II HK and Ca II NIR, respectively.}
\end{figure*}

Some examples of the time evolution of FWHM are shown in Fig.\ref{Fig1} (data and uncertainties are presented in Tab.\ref{Tab1}). The example in the upper left panel is SN 2011fe, a typical `NV' subtype. One can see that the evolution is very slow, especially after -8 days. Exceptions are those broad lines at early phases, for example Si II $\lambda$6355 and Ca II NIR at $t<-6$ days. A possible reason is the saturation \citep{Zhao21}, which can deform the absorption lines, increasing the measured FWHM. Another possible reason is the velocity, which decreases with time at early phases. FWHM may decrease accordingly, for Doppler-broadening's dependence on the velocity.

Shown in the upper right panel of Fig.\ref{Fig1} is a typical `HV' subtype SN Ia, the SN 2002bo. At early phases, the FWHM of Si II $\lambda$ 6355 ($\gamma^{Si6}$) of this object is comparable to that of SN 2011fe (e.g. $\approx 220$ \AA~at -12 days), but declines much slower with time. At phases near maximum light, $\gamma^{Si6}$ of SN 2002bo becomes much larger than that of SN 2011fe ( $\approx$ 200 ~vs. 150 \AA), possibly due to a higher abundance of the ion (note that the effect is less significant at early phases when the line is more saturated). Shown in the bottom left panel is the 1991T/1999aa-like object SN 1999aa. It shows a sharp decreasing of the FWHMs of Si II $\lambda$s 5972 and 6355. At -10 days, they reach a maximum of nearly 300 \AA, while 6 days latter, at -4 days they drop to only about 150 \AA. Also note that, at -11 days, the FWHM of Ca II NIR is relatively small. A direct reason is very weak absorption (pEW<10\AA), possibly due to too high temperature. Finally, an example of 1991bg-like object, the SN 1999by, is shown in the bottom right panel. As one can see, the FWHM of Si II $\lambda$ 6355 increases with time at phases near maximum light. Though the reason is unclear, we do see similar trend of the velocity in our previous work \citet{Zhao21}.

\section{Correlation with rest wavelength\label{S4}}

\begin{figure*}
	\centering
	\includegraphics[width=1\columnwidth]{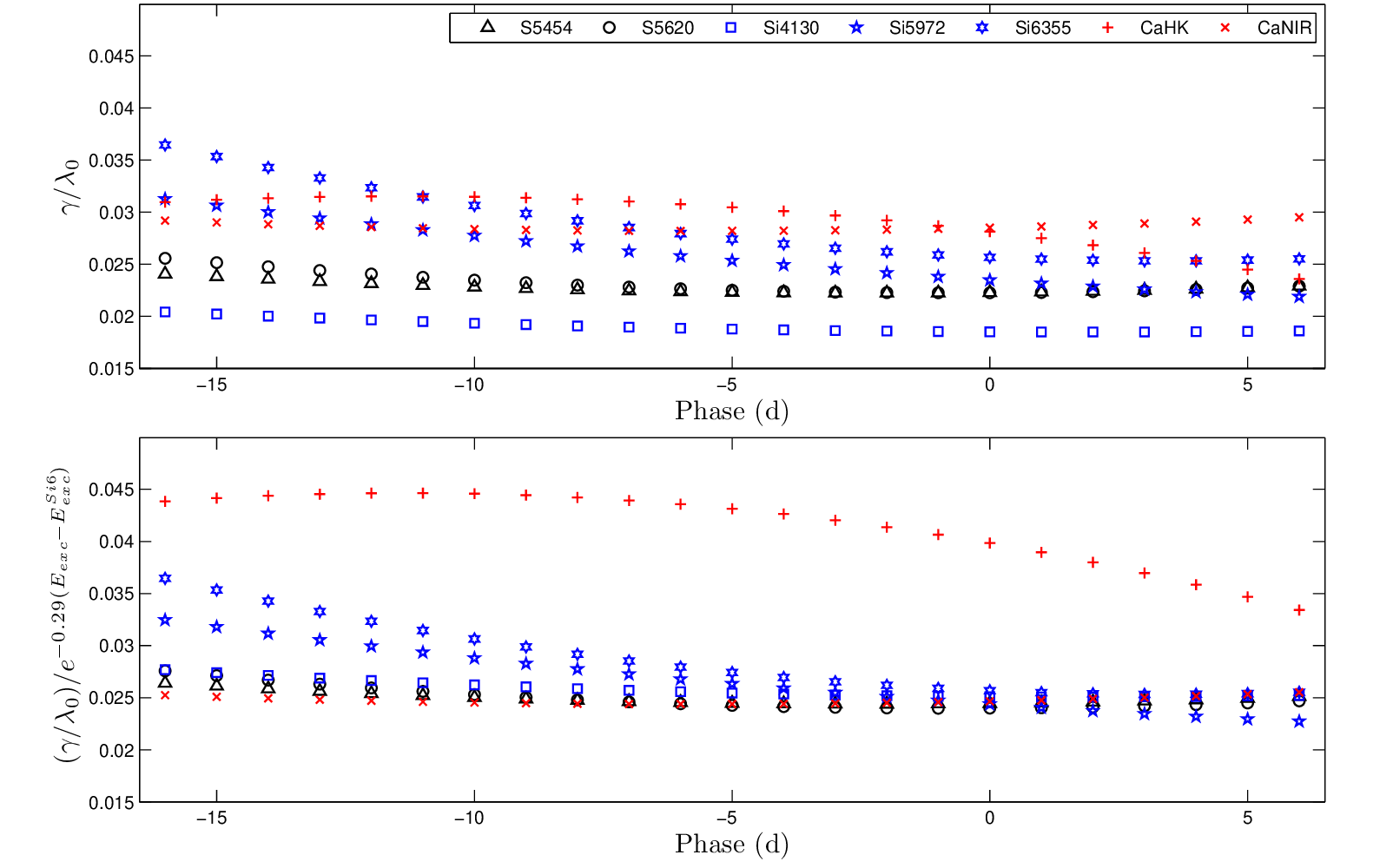}
	\caption{\label{Fig2} Shown in the upper panel are fitting results of the ratios of FWHMs and wavelengths ($R=\gamma/\lambda_0$) for several lines. Data are mostly from the sample presented in \citet{Zhao21, Zhao24} and fitted with second-order polynomials. But, the sample size is too large, and we are only interested in the overall trend, so they are not fully presented in Tab.\ref{Tab1} \& \ref{Tab2}. Shown below is also the ratio, but scaled by a factor of $e^{-0.29(E_{exc}-E_{exc}^{Si6})}$, where `$E_{exc}$' is the line's excitation energy (i.e. $\frac{hc}{\lambda_0}$), and `$E_{exc}^{Si6}$' is that of line Si II $\lambda$6355. }
\end{figure*}

Fig.\ref{Fig2} shows how FWHM correlates with the rest-wavelength. Rest-wavelength is 3945, 4130, 5454\&5620, 5972, 6355, 8567 \AA~ for Ca II HK, Si II $\lambda$4130, S II W-trough, Si II $\lambda$5972, Si II $\lambda$6355 and Ca II NIR, respectively. It is clear that shorter lines have a narrower FWHM. Therefore, a speculation is that wavelength may play an important role. In fact, the widths of absorption features are mainly decided by Doppler-broadening, which originated from kinetic motions, including thermal motion and turbulence. A rough expression is $\Delta \lambda \approx \frac{\Delta V}{c}\lambda_0$, where `$V$' is the velocity, `$c$' is the speed of light. Obviously, it is proportional to the rest-wavelength $\lambda_0$. 

Since the rest wavelength is decided by the excitation energy, i.e. $E_{exc}=\frac{hc}{\lambda}$, in essence FWHM is dependent on the excitation energy. And, since $pEW\approx 1.064 \gamma$ (for Gaussian profiles), the line strengths should also depend on the excitation energy. This, actually, has been confirmed by previous measurement results \citep[see for example,][]{Zhao21,Zhao24}. 

In Fig.\ref{Fig2}, the ratio of FWHM to wavelength, $R=\gamma/\lambda_0$, is shown as a function of the time. Generally speaking, the ratio is close to 0.015 for all lines, and barely evolve with time after -6 days. Median errors (which is less affected by outliers) are $\approx$ 0.0007, 0.0006, 0.0017, 0.0016, 0.0017, 0.0018, 0.0005, for lines S II $\lambda$5454, S II $\lambda$5620, Si II $\lambda$4130, Si II $\lambda$5972, Si II $\lambda$6355, Ca II HK, and Ca II NIR, respectively. In other words, the error level is mostly below 10\%.

In the below panel of Fig.\ref{Fig2}, we show a scaled version of the ratio $R$. The scaling factor is expressed as an exponential function $e^{-0.29(E_{exc}-E_{exc}^{Si6})}$. This further reduces their differences, especially after -5 days, except for line Ca II HK. The reason is unclear, though this line does have some specificity, like almost zero lower level, and very short wavelength $\lambda\approx3945$ \AA. 

The relation between FWHM and the wavelength can be helpful in spectral measurements. Usually, the velocities of the lines are easier to determined, while FWHM, or $\sigma$ in the Gaussian function is harder to determine, especially when line blending is serious. An example is the fitting of Si II $\lambda$5972. This line is relatively weak, and to some extend affected by HVF of Si II $\lambda$6355 (or some unknown absorption that lies between Si II $\lambda$5972 and Si II $\lambda$6355). This often makes it difficult to accurately determine the FWHM of Si II $\lambda$5972. In comparison, FWHM of Si II $\lambda$6355 is often much easier to determine, if its HVF is not too strong. Therefore, we may use the FWHM of Si II $\lambda$6355 to help determine that of Si II $\lambda$5972. 


\section{Diversity among different subclasses\label{S5}}

\begin{figure}
	\includegraphics[width=1\columnwidth]{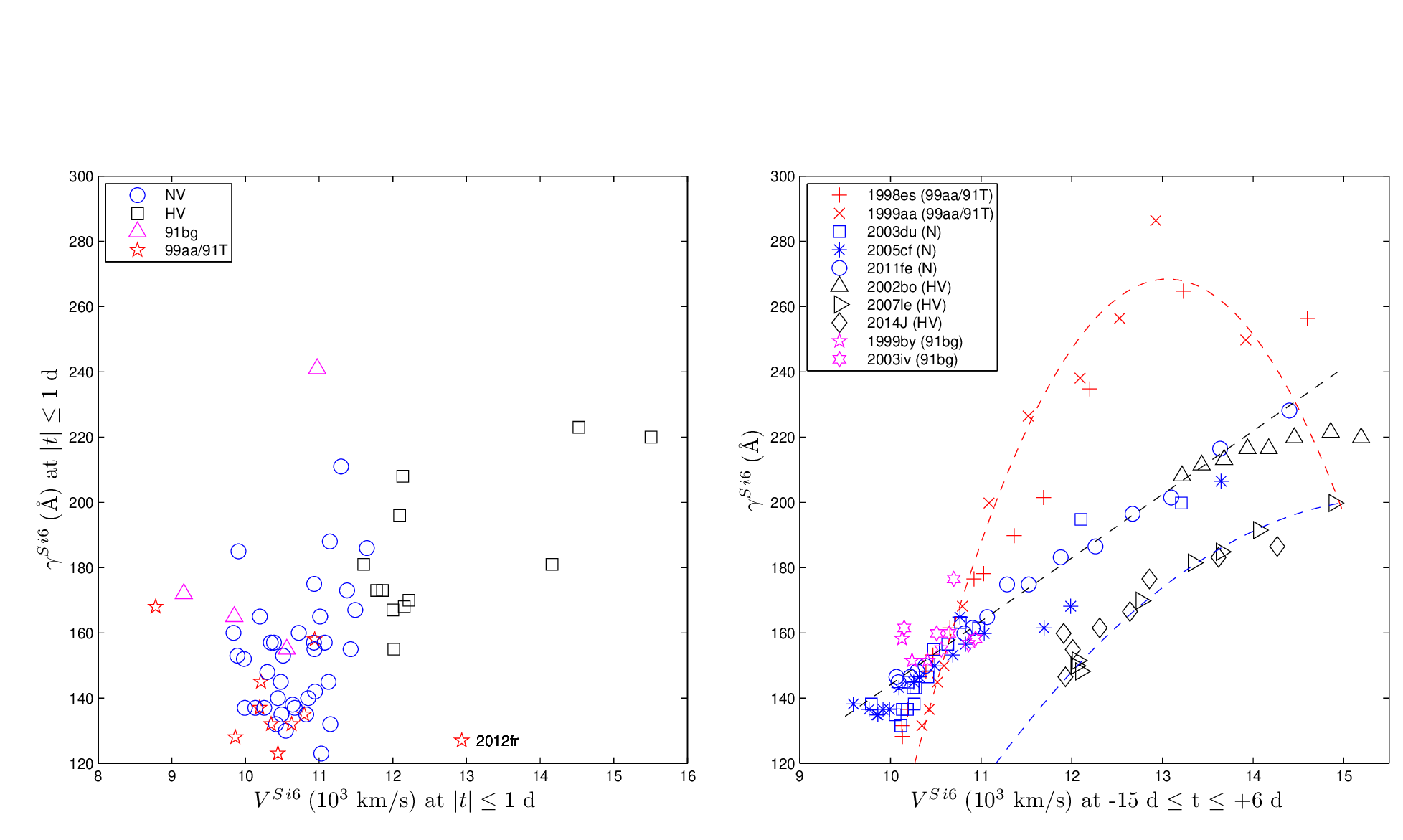} 
	\caption{\label{Fig3}. Correlation between the velocity and FWHM of Si II $\lambda$ 6355. Left panel: A sample at phases near maximum light, i.e. $|t|\leq 1$ days. Each point represents an individual object. Right panel: Some typical objects, at phases $\leq$ +6 days. Classifications are according to \citet{Wang09}.}
\end{figure}

As mentioned before, diversity of SN Ia can be roughly indicated by the velocity. Usually, those with relatively higher velocities also have larger line strengths \citep{Wang09}. Physically, the line strength represents absorption rate of the photons, which is mainly determined by the abundance of the absorbing ions and the energy levels of their electrons. Therefore, HV SNe may have more abundant IMEs in their ejecta. Now, a key question is whether higher velocity also results in a larger FWHM. The results of our investigation is shown in the left panel of Figure \ref{Fig3} (data are presented in Tab.\ref{Tab2}). Apparently, indeed HV objects have relatively larger FWHM. This result is expected, as Doppler broadening depends on the ejecta's velocity.

In the right panel, we show a few typical objects. One can see that for each object, FWHM basically increases with the velocity. This is expected, as Doppler broadening (which is determined by velocity distributions in the ejecta) is the main origin of the FWHM. However, if the velocity is the same, then it appears that NV object tends to have larger FWHM than HV objects. A possible reason is that NV objects would then typically appears at earlier phases, when the temperatures could be relatively higher. An additional evidence is that 1991T/1999aa-like objects show the largest FWHM. At first glance, these might seem opposite to those in the left panel, but in fact, they only reflect the fact that other factors (other than the absolute value of velocity) also affect the FWHM. It should also be noted that here the comparison is not necessarily at the same phase. For NV objects in the right panel, the correlation is fitted as $\gamma\approx 0.02V-50$, where `$V$' is the velocity (km/s). Cold 1991bg-like objects show low velocity and also low FWHM. 

\begin{figure}
	\includegraphics[width=1\columnwidth]{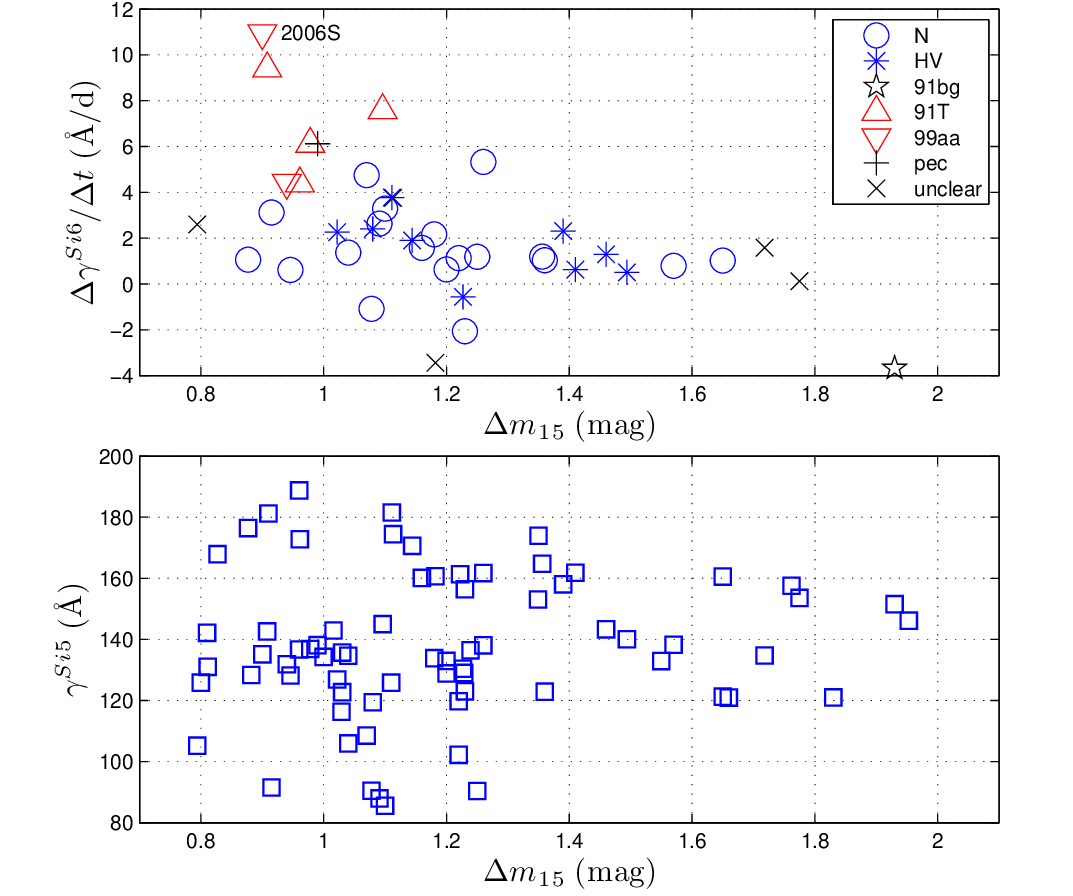}
	\caption{\label{Fig4} Upper panel: decline rate of FWHM of Si II $\lambda$6355 ($\Delta\gamma^{Si6}/\Delta t$) as a function of the decline rate of brightness ($\Delta m_{15}$). Here $\Delta\gamma^{Si6}$ is calculated by $\Delta\gamma^{Si6}=\gamma^{Si6}_{t1}-\gamma^{Si6}_{t2}$, with $t_1\approx -4$ days and $t_2\approx 0$ days. $\Delta t$ is calculated by $\Delta t=t_2-t_1$. Lower panel: $\Delta m_{15}$ vs. the FWHM of Si II $\lambda$5972 of SNe Ia at the time of B-band maximum light.}
\end{figure}

A four-days change of FWHM is shown in the upper panel of Fig.\ref{Fig4}, as a function of the decline rate (data are from Tab.\ref{Tab2}). Apparently, 1991T/1999aa-like objects show much greater variation than other objects. This is understandable, as higher explosion temperature is expected to cause greater Doppler broadening than colder objects. This regulation can help identify the subtype. An example is SN 2006S which is easily mistaken for a NV object. As shown in the upper panels of Fig.\ref{Fig4}, this object actually has a very rapid variation of FWHM, which is a sign of 1991T/1999aa-like object. A more direct evidence is, of course, its relatively high luminosity ($\Delta m_{15}\approx 0.9$ mag). A similar case is the outlier SN 2012fr in the left panel of Fig.\ref{Fig3}, which also shows a rapid decrease of the FWHM of Si II $\lambda$6355 (an approximating polynomial of degree 2 suggests $\gamma^{Si6}\approx 193$ \AA~at -4 days, and $\approx$ 168 \AA~at 0 days, so $\Delta\gamma^{Si6}/\Delta t\approx 6.3$ \AA/d) and a very bright luminosity ($\Delta m_{15}\approx 0.8$ mag). So, we also consider it a 1991T/1999aa-like object, rather than `HV'. More detailed observations and profound analysis of this object can be found in e.g. \citet{Childress13}; \citet{Zhangjj14}. In the lower panel of Fig.\ref{Fig4}, the FWHM of Si II $\lambda$5972 at maximum light is plotted against the decline rate. No clear correlation is seen, meaning the FWHM is possibly independent of the decline rate, or only insignificantly affected. 




\section{Possible applications of the FWHM \label{S6}}

\begin{figure}
	\includegraphics[width=1\columnwidth]{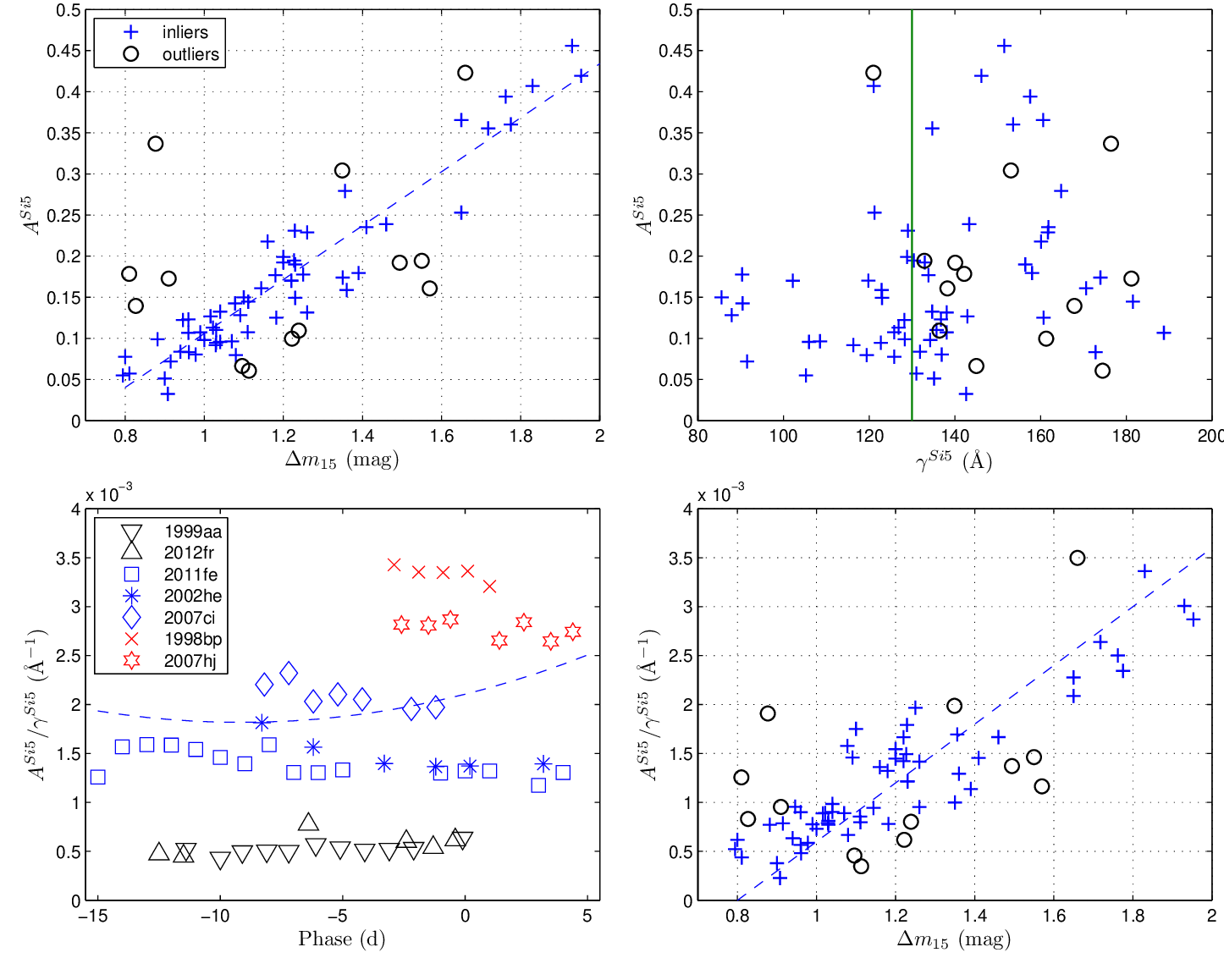}
	\caption{\label{Fig5} Upper left panel shows the absorption depth of Si II $\lambda$5972 ($A^{Si5}$) as a function of the decline rate ($\Delta m_{15}$). Plus signs mark the inliers which have relatively small residuals from the fit, i.e. $\Delta A^{Si5}=|A^{Si5}-A^{Si5}_{fitted}|\leq 0.07$, where $A^{Si5}_{fitted}$ is given by the best-fit linear relation. Black circles mark the outliers which have relatively large residuals, i.e. $\Delta A^{Si5}=|A^{Si5}-A^{Si5}_{fitted}|> 0.07$. All objects are measured at $|t|\leq 1$ days, each point represents an individual object. The straight dash line in this panel is only a guiding line, not a fitting line. Upper right panel shows $A^{Si5}$ as a function of the FWHM ($\gamma^{Si5}$). Markers in this panel and the lower right panel are the same as for the upper left panel. Lower left panel shows the temporal evolution of the ratio of depth to FWHM of line Si II $\lambda$5972 ($A^{Si5}/\gamma^{Si5}$). The dash line in this panel is a fitting line with a large sample (data not completely presented in this paper, for the same reason as in Fig.\ref{Fig2}). Lower right panel shows the correlation between $A^{Si5}/\gamma^{Si5}$ and the decline rate. The straight dash line in this panel is also a guiding lines.}
\end{figure}

In most studies, the absorption strength was represented with pEW. But in fact, its correlation with $\Delta m_{15}$ is weaker than the maximum absorption depth. A possible reason is that, pEW is also affected by FWHM which has a relatively high uncertainty (see Section \ref{S2}). As mentioned before, the most sensitive indicator of the decline rate may be $A^{Si5}$. An example is shown in the upper left panel of Figure \ref{Fig5} (phases are at $|t|\leq 1$ days; data are presented in Tab.\ref{Tab3}). Based on the residuals from the fit, the sample is split into two groups: the inliers with $\Delta A^{Si5}=|A^{Si5}-A^{Si5}_{fitted}|\leq 0.07$ and the outliers with $\Delta A^{Si5} > 0.07$. As shown in the upper right panel, the outliers are predominantly found at $\gamma^{Si5}>$ 130 \AA. This means that objects with $\gamma^{Si5} \leq 130$~\AA ~are predominantly inliers in the $A^{Si5}$--$\Delta m_{15}$ correlation. Also, they appear to span a narrower range of $A^{Si5}$, typically from 0.05 to 0.25. In comparison, objects with $\gamma^{Si5} > 130~\text{\AA}$ span a broader range of $A^{Si5}$, typically from 0.05 to 0.40. These results may imply relatively low diversity among these objects with $\gamma^{Si5} \leq 130$~\AA. The reason is unclear, but a more symmetry explosion, with less collisions, may lead to less velocity dispersion and so a smaller FWHM. Also, interfering factors like saturation may be weaker in these objects. Anyway, if they are truly a more homogeneous sample, then one may consider using such a smaller sample to calibrate the cosmic distance.

Many spectral parameters have been found to have a strong correlation with the decline rate, which can be used to estimate the brightness. But this, in fact, is not practicable, because usually the spectral parameters significantly vary with time. And, determining the phase itself requires light curve, the lack of which is the very reason that we have to use a spectroscopic method. For example, the `Si II ratio' $R_{si}=A^{Si5}/A^{Si6}$ as defined by \citep{Nugent95} also has a strong correlation with $\Delta m_{15}$ (Pearson's correlation coefficient $\rho=0.918$ for `good' candidates in Fig.\ref{Fig5})), but its time evolution is very significant (mostly due to the evolution of Si II $\lambda$6355). $A_{Si5}$ has the strongest correlation with $\Delta m_{15}$ ($\rho=0.951$), and also one of the slowest evolution, but the error level is still too high. The best choice may be the ratio of depth to FWHM of Si II $\lambda$5972 ($R_{A\gamma}^{Si5}=A^{Si5}/\gamma^{Si5}$). This ratio has a strong correlation with $\Delta m_{15}$ ($\rho=0.910$), and luckily, also a very slow temporal evolution, as shown in the lower left panel of Fig.\ref{Fig5}. The ratio can be considered as a shape index (the larger it is, the sharper the curve is), so its quasi-invariance suggests that the absorption features vary in a self-similar style. Average value of $R_{A\gamma}^{Si5}$ is near 0.002 \AA. The increase near maximum light may be partially due to increase of dimmer objects in the sample. Assuming that the spectrum is obtained at phases between -15 days to +5 days, a rough estimation is $\Delta m_{15}\approx 0.725+373A^{Si5}/\gamma^{Si5}$. The formula is fitted from those `good candidates' in the lower-right panel of Fig.\ref{Fig5}. A test with 883 spectra yields median absolute errors (i.e. 50\%-confidence uncertainty, here we did not use standard deviation, for too large influence from outliers) $\widetilde{E}\lesssim$ 0.20 mag at $t \leq -10$ days, and $\widetilde{E}\lesssim$ 0.11 mag at $-10 \leq t \leq +3$ days. A further improvement is still possible, if line Si II $\lambda$5972 is further studied, or taking into consideration other factors like the subtype, but this is beyond the scope of this study.

\section{Conclusion}

Spectral features of SNe Ia carry important information about the explosion. But unfortunately, they are difficult to interpret, as multiple factors affect. In this study, we focuses on an important one, i.e. the FWHM of the absorption features. It is a crucial parameter for spectral fitting. And, together with absorption depth, it decides the line strength. 

Our results suggest that, FWHM is mainly determined by the wavelength of the line, or equivalently the excitation energy. Difference in FWHM can be significantly reduced by dividing their wavelengths. The reason may be Doppler broadening, which is roughly expressed as $\Delta \lambda \approx \frac{\Delta v}{c}\lambda_0$. We also found that, further scaling with a factor $e^{-0.29(E_{exc}-E_{exc}^{Si6})}$ can further reduce the difference of FWHM among lines Si II $\lambda\lambda$ 4130, 5972, 6355, Ca II NIR, and the two lines of S II W-trough. The only exception is Ca II HK which itself is special for having almost zero $E_{lower}$ and the shortest wavelength. This correlation is helpful in fitting lines that are seriously blended, for example the important line Si II $\lambda$5972.

Besides wavelength, FWHM is also affected by velocity and possibly the temperature. Generally speaking, higher velocity often corresponds to larger FWHM. And, higher temperature may be the reason that 1991T/1999aa-like objects show the largest FWHM, if compared at the same velocity. 1991T/1999aa-like objects also show much faster time evolution of FWHM, which may be used to help identify the subtype.

We found that, objects with FWHM of Si II $\lambda$5972 smaller than 130 \AA~have much tighter correlation between the absorption depth of Si II $\lambda$5972 and $\Delta m_{15}$. This may suggest a new constrain on the sample to further improve the accuracy of distance's measurement in cosmology. 

We also found that Si II $\lambda$5972 happens to have a relatively slow time evolution. So, in case when light curve is not available, a very rough estimation can be obtained from relation $\Delta m_{15}\approx 0.725+373A^{Si5}/\gamma^{Si5}$ (assuming that the phase is between -15 and +5 days).


\section*{Acknowledgements}
We thank the anonymous referee, for comments which significantly elevated the level of this paper. This research has made use of the CfA Supernova Archive, which is funded in part by the National Science Foundation through grant AST 0907903, the Berkeley/Lick Supernova Archives, which is funded in part by the US National Science Foundation, and the CSP Supernova Archive, which is supported by the World Premier International Research Center Initiative. Here we express our deep gratitude.

\clearpage


\begin{longtable}{ccccccccccccccc}
	\caption{Temporal evolution of the FWHMs of several important absorption features of SN 2011fe \label{Tab1}}\\
	\hline
	SN & Phase & $\gamma^{Si4}$ & $\gamma^{Si5}$ & $\gamma^{Si6}$ & $\gamma^{S1}$ & $\gamma^{S2}$ & $\gamma^{CaHK}$ & $\gamma^{CaNIR}$ \\
	~ &(days) & (\AA) & (\AA) & (\AA) & (\AA) & (\AA) & (\AA) & (\AA) \\
    \endfirsthead
	\hline

1999aa & -- & -- & 233(11) & 251(4) & 132(7) & 140(4) & 133(7) & -- \\
1999aa & -- & -- & -- & -- & -- & -- & -- & 149(11) \\
1999aa & -10 & 50(3) & 286(15) & 286(3) & 130(7) & 139(4) & 133(7) & -- \\
1999aa & -9.1 & 50(3) & 257(13) & 257(2) & 125(7) & 134(2) & 133(7) & -- \\
1999aa & -8.1 & 50(3) & 238(12) & 238(2) & 120(15) & 128(12) & 133(7) & -- \\
1999aa & -7.2 & 54(3) & 226(12) & 226(3) & 117(6) & 125(2) & 133(7) & -- \\
1999aa & -6.1 & 58(3) & 199(11) & 199(2) & 117(6) & 126(2) & 133(7) & -- \\
1999aa & -5.1 & 57(2) & 168(9) & 168(9) & 113(6) & 122(7) & 133(7) & -- \\
1999aa & -4.1 & 53(2) & 150(8) & 150(8) & 118(6) & 126(7) & -- & -- \\
1999aa & -3.1 & 50(3) & 145(8) & 145(8) & 119(6) & 127(7) & 133(7) & -- \\
1999aa & -2.1 & 50(3) & 137(7) & 137(7) & 124(7) & 132(7) & -- & -- \\
1999aa & -- & -- & -- & -- & -- & -- & 131(4) & -- \\
1999aa & -0.1 & 50(3) & 132(7) & 132(7) & 139(7) & 131(7) & -- & -- \\
1999aa & -- & -- & -- & -- & -- & -- & -- & 203(6) \\
1999by & -6.2 & 107(6) & 125(6) & 152(2) & 122(7) & 130(4) & -- & -- \\
1999by & -5.2 & 102(6) & 122(4) & 152(2) & 120(7) & 129(2) & -- & -- \\
1999by & -4.2 & 97(6) & 131(4) & 158(2) & 124(7) & 133(2) & 147(8) & -- \\
1999by & -- & -- & -- & -- & -- & -- & -- & 333(18) \\
1999by & -3.2 & 93(6) & -- & -- & -- & -- & -- & -- \\
1999by & -- & -- & -- & -- & -- & -- & 146(8) & 333(18) \\
1999by & 1.8 & 100(9) & 121(7) & 192(10) & 92(5) & 101(6) & 135(8) & -- \\
1999by & -- & -- & -- & -- & -- & -- & -- & 332(12) \\
1999by & 2.8 & 103(9) & 118(6) & 202(11) & 79(4) & 87(5) & -- & -- \\
1999by & -- & -- & -- & -- & -- & -- & -- & 325(11) \\
1999by & 3.8 & 92(9) & -- & -- & -- & -- & -- & -- \\
1999by & 4.8 & 89(9) & -- & -- & -- & -- & -- & -- \\
2002bo & -13.4 & 101(4) & -- & -- & -- & -- & -- & -- \\
2002bo & -- & -- & -- & -- & -- & -- & -- & 242(6) \\
2002bo & -12.5 & 84(4) & 181(10) & 221(12) & 144(8) & 152(8) & -- & -- \\
2002bo & -8.4 & 76(2) & 155(8) & 220(12) & 126(7) & 135(7) & -- & -- \\
2002bo & -7.4 & 75(6) & 142(8) & 222(12) & 123(7) & 131(7) & -- & -- \\
2002bo & -6.3 & 77(2) & 139(7) & 220(12) & 121(7) & 130(7) & -- & -- \\
2002bo & -5.4 & 93(2) & 147(8) & 216(11) & 122(7) & 130(7) & -- & -- \\
2002bo & -4.4 & 91(2) & 149(8) & 216(11) & 119(6) & 128(7) & -- & -- \\
2002bo & -- & -- & -- & -- & -- & -- & -- & 278(4) \\
2002bo & -3.4 & 95(2) & 143(8) & 214(11) & 121(7) & 129(7) & -- & -- \\
2002bo & -- & -- & -- & -- & -- & -- & -- & 262(4) \\
2002bo & -2.5 & 92(2) & 147(8) & 212(11) & 121(7) & 129(7) & -- & -- \\
2002bo & -- & -- & -- & -- & -- & -- & -- & 281(6) \\
2002bo & -1.5 & 97(2) & 150(8) & 208(11) & 120(7) & 128(7) & -- & -- \\
2002bo & -- & -- & -- & -- & -- & -- & -- & 252(6) \\
2011fe & -- & -- & -- & -- & -- & -- & -- & 280(3) \\
2011fe & -15 & 84(5) & 192(7) & 228(2) & 99(6) & 108(4) & 129(7) & 307(2) \\
2011fe & -14 & 85(4) & 187(6) & 217(2) & 119(6) & 127(4) & 126(7) & 346(2) \\
2011fe & -13 & 88(4) & 179(4) & 202(2) & 127(7) & 135(2) & 128(7) & 353(4) \\
2011fe & -12 & 77(4) & 185(4) & 196(2) & 126(7) & 134(2) & 117(6) & 342(6) \\
2011fe & -11 & 72(4) & 181(4) & 187(2) & 123(7) & 131(2) & 115(6) & 335(4) \\
2011fe & -10 & 79(4) & 182(4) & 183(2) & 121(7) & 129(2) & 121(7) & 317(4) \\
2011fe & -9 & 77(4) & 175(6) & 175(2) & 119(6) & 127(2) & 119(6) & 280(4) \\
2011fe & -8 & 83(5) & 150(8) & 175(2) & 119(6) & 127(2) & 104(2) & 268(4) \\
2011fe & -7 & 77(4) & 165(9) & 165(2) & 117(6) & 126(2) & 114(4) & 260(4) \\
2011fe & -6 & 77(4) & 161(9) & 161(2) & 116(4) & 124(4) & 110(4) & 261(4) \\
2011fe & -5 & 72(2) & 159(8) & 159(2) & 120(6) & 128(4) & 115(4) & 248(4) \\
2011fe & -- & -- & -- & -- & -- & -- & 118(2) & 252(4) \\
2011fe & -- & -- & -- & -- & -- & -- & -- & 266(14) \\
2011fe & -1 & 67(2) & 135(6) & 149(2) & 113(4) & 119(2) & 112(2) & 250(2) \\
2011fe & 0 & 68(2) & 134(7) & 149(3) & 114(4) & 115(2) & 103(2) & 252(2) \\
2011fe & 1 & 65(2) & 131(7) & 147(8) & 114(6) & 114(6) & 105(2) & 234(4) \\
2011fe & 2 & 66(2) & -- & -- & -- & -- & 100(2) & 237(4) \\
2011fe & 3 & 66(2) & 143(8) & 145(8) & 112(6) & 114(6) & 103(6) & 240(4) \\
2011fe & 4 & 71(2) & 147(8) & 147(8) & 113(6) & 115(6) & 105(6) & 247(4) \\

\hline
\end{longtable}

Note: Data sources and photometric parameters can be found in \citet{Zhao21,Zhao24}. Phases are from the $B-$band maximum light.	$\gamma$ represents the full with half maximum (FWHM). Superscripts `$Si4$', `$Si5$', `$Si6$', `$S1$', `$S2$', `$CaHK$' and `$CaNIR$' mark lines Si II $\lambda$4130, Si II $\lambda$5972, Si II $\lambda$6355, S II $\lambda$5454, S II $\lambda$5620, Ca II HK and Ca II NIR, respectively. \\

\setlength{\tabcolsep}{4pt}
\begin{longtable}{cccccc|cccccc}
	\caption{The velocities and FWHMs of Si II $\lambda$6355 \label{Tab2}}\\
	
	\hline
	SN & Phase & $V^{Si6}$ & $\gamma^{Si6}$ & subtype & $\Delta m_{15}$ & SN & Phase & $V^{Si6}$ & $\gamma^{Si6}$ & subtype & $\Delta m_{15}$ \\
	~ & (days) & (km/s) & (\AA) & ~ & (mag) & ~ & (days) & (km/s) & (\AA) & ~ & (mag) \\
	\hline
	\endfirsthead
	
\multicolumn{4}{l}{\textbf{Table 2 continued}} \\
\hline
SN & Phase & $V^{Si6}$ & $\gamma^{Si6}$ & subtype & $\Delta m_{15}$ & SN & Phase & $V^{Si6}$ & $\gamma^{Si6}$ & subtype & $\Delta m_{15}$ \\
~ & (days) & (km/s) & (\AA) & ~ & (mag) & ~ & (days) & (km/s) & (\AA) & ~ & (mag) \\
\hline
\endhead
	
\hline
\multicolumn{4}{l}{Continued on next page} \\
\endfoot	

\endlastfoot
	
1986G & -4 & 11555(127) & 188(10) & NV & 1.65 & 2003du & -0.1 & 10133(119) & 136(7) & NV & 1.07 \\ 
1986G & 0 & 9904(109) & 184(10) & NV & 1.65 & 2003it & -3.3 & 11734(86) & 158(3) & NV & 1.36 \\ 
1992A & -4 & 12579(122) & 176(3) & HV & 1.41 & 2003it & -0.2 & 11428(67) & 155(2) & NV & 1.36 \\ 
1992A & 0 & 11860(170) & 173(9) & HV & 1.41 & 2003iv & -2.5 & 10938(215) & 158(4) & 91bg & 1.65 \\ 
1995E & -4.2 & 11118(142) & 166(9) & NV & 1.16 & 2003iv & -1.9 & 10862(163) & 157(9) & 91bg & 1.65 \\ 
1995E & -0.2 & 10722(135) & 160(9) & NV & 1.16 & 2003iv & -1.5 & 10862(163) & 157(9) & 91bg & 1.65 \\ 
1996X & -4.1 & 11188(65) & 160(2) & NV & 1.26 & 2003iv & -0.9 & 10649(153) & 160(9) & 91bg & 1.65 \\ 
1996X & -0.1 & 10640(107) & 138(3) & NV & 1.26 & 2003iv & -0.5 & 10649(153) & 160(9) & 91bg & 1.65 \\ 
1997bp & -3.1 & 16000(174) & 226(12) & HV & 1.08 & 2003iv & 0.1 & 10561(163) & 155(3) & 91bg & 1.65 \\ 
1997bp & -0.2 & 15506(167) & 219(11) & HV & 1.08 & 2003iv & 0.5 & 10561(163) & 155(3) & 91bg & 1.65 \\ 
1997dt & -4.8 & 11649(43) & 163(2) & NV & 1.04 & 2003iv & 1.1 & 10510(131) & 160(3) & 91bg & 1.65 \\ 
1997dt & 0.2 & 10929(118) & 156(8) & NV & 1.04 & 2003iv & 1.5 & 10510(131) & 160(3) & 91bg & 1.65 \\ 
1998de & -3.2 & 11865(208) & 230(5) & 91bg & 1.93 & 2003iv & 1.6 & 10699(130) & 177(9) & 91bg & 1.65 \\ 
1998de & -0.3 & 10973(160) & 241(13) & 91bg & 1.93 & 2003iv & 3.5 & 10152(139) & 162(3) & 91bg & 1.65 \\ 
1998dh & -3.5 & 12000(131) & 165(9) & HV & 1.227 & 2004as & -4 & 12797(170) & 212(11) & HV & 1.111 \\ 
1998dh & -0.5 & 12000(131) & 166(9) & HV & 1.227 & 2004as & -0.1 & 12092(152) & 197(10) & HV & 1.111 \\ 
1998es & -11.5 & 14598(90) & 256(3) & 91T & 0.978 & 2004at & -3.7 & 10978(127) & 155(8) & NV & 1.091 \\ 
1998es & -10.4 & 13231(84) & 265(3) & 91T & 0.978 & 2004at & 0.2 & 10480(53) & 145(2) & NV & 1.091 \\ 
1998es & -9.4 & 12200(129) & 235(2) & 91T & 0.978 & 2004gs & -3.8 & 11678(178) & 196(11) & unclear & 1.775 \\ 
1998es & -8.5 & 11689(60) & 201(4) & 91T & 0.978 & 2004gs & -0.3 & 10893(168) & 195(11) & unclear & 1.775 \\ 
1998es & -7.4 & 11365(118) & 190(10) & 91T & 0.978 & 2005cf & -12.7 & 13648(65) & 206(2) & NV & 1.1 \\ 
1998es & -6.4 & 11028(117) & 178(9) & 91T & 0.978 & 2005cf & -10.7 & 11989(129) & 168(9) & NV & 1.1 \\ 
1998es & -5.4 & 10921(115) & 177(9) & 91T & 0.978 & 2005cf & -9.7 & 11695(127) & 162(9) & NV & 1.1 \\ 
1998es & -4.5 & 10656(112) & 162(9) & 91T & 0.978 & 2005cf & -7.7 & 11033(126) & 160(9) & NV & 1.1 \\ 
1998es & -3.4 & 10468(39) & 153(2) & 91T & 0.978 & 2005cf & -6.7 & 10767(94) & 165(3) & NV & 1.1 \\ 
1998es & -2.4 & 10392(112) & 148(8) & 91T & 0.978 & 2005cf & -5.7 & 10829(58) & 157(2) & NV & 1.1 \\ 
1998es & -0.5 & 10190(109) & 137(7) & 91T & 0.978 & 2005cf & -4.7 & 10689(59) & 153(2) & NV & 1.1 \\ 
1998es & 0.5 & 10126(107) & 132(7) & 91T & 0.978 & 2005cf & -3.7 & 10488(61) & 150(2) & NV & 1.1 \\ 
1998es & 1.5 & 10133(110) & 128(7) & 91T & 0.978 & 2005cf & -2.7 & 10320(56) & 147(2) & NV & 1.1 \\ 
1998es & -4.5 & 10656(112) & 161(9) & 91T & 0.978 & 2005cf & -1.7 & 10257(52) & 145(2) & NV & 1.1 \\ 
1998es & -0.5 & 10190(109) & 137(7) & 91T & 0.978 & 2005cf & -0.7 & 10094(60) & 143(2) & NV & 1.1 \\ 
1999aa & -11.4 & 13919(116) & 250(4) & 99aa & 0.94 & 2005cf & 0.3 & 9990(54) & 137(2) & NV & 1.1 \\ 
1999aa & -10 & 12927(72) & 286(3) & 99aa & 0.94 & 2005cf & 1.3 & 9918(53) & 137(2) & NV & 1.1 \\ 
1999aa & -9.1 & 12528(57) & 256(2) & 99aa & 0.94 & 2005cf & 2.3 & 9852(112) & 135(7) & NV & 1.1 \\ 
1999aa & -8.1 & 12092(57) & 238(2) & 99aa & 0.94 & 2005cf & 3.3 & 9860(111) & 135(7) & NV & 1.1 \\ 
1999aa & -7.2 & 11520(68) & 226(3) & 99aa & 0.94 & 2005cf & 4.3 & 9765(117) & 137(7) & NV & 1.1 \\ 
1999aa & -6.1 & 11087(65) & 200(2) & 99aa & 0.94 & 2005cf & 5.3 & 9592(119) & 138(7) & NV & 1.1 \\ 
1999aa & -5.1 & 10795(116) & 168(9) & 99aa & 0.94 & 2005cf & -3.7 & 10488(61) & 150(2) & NV & 1.1 \\ 
1999aa & -4.1 & 10592(112) & 150(8) & 99aa & 0.94 & 2005cf & 0.3 & 9990(54) & 137(2) & NV & 1.1 \\ 
1999aa & -3.1 & 10518(111) & 145(8) & 99aa & 0.94 & 2006N & -3.4 & 11474(125) & 158(8) & NV & 1.57 \\ 
1999aa & -2.1 & 10428(111) & 137(7) & 99aa & 0.94 & 2006N & 0.7 & 10935(123) & 155(8) & NV & 1.57 \\ 
1999aa & -0.1 & 10348(112) & 132(7) & 99aa & 0.94 & 2006S & -4.7 & 11805(144) & 190(10) & 99aa & 0.9 \\ 
1999aa & -4.1 & 10592(112) & 150(8) & 99aa & 0.94 & 2006S & 0.3 & 10797(128) & 135(7) & 99aa & 0.9 \\ 
1999aa & -0.1 & 10348(112) & 132(7) & 99aa & 0.94 & 2006bt & -4 & 12224(141) & 216(11) & NV & 0.877 \\ 
1999ac & -4 & 11070(36) & 163(2) & unclear & 1.182 & 2006bt & -0.1 & 11298(157) & 212(11) & NV & 0.877 \\ 
1999ac & -1 & 10168(40) & 173(2) & unclear & 1.182 & 2006sr & -3 & 12477(137) & 175(9) & HV & 1.39 \\ 
1999by & -6.2 & 10408(48) & 152(2) & 91bg & 1.899 & 2006sr & -1 & 12218(143) & 170(9) & HV & 1.39 \\ 
1999by & -5.2 & 10239(58) & 152(2) & 91bg & 1.899 & 2007A & -3.5 & 10791(76) & 142(2) & NV & 0.946 \\ 
1999by & -4.2 & 10128(61) & 158(2) & 91bg & 1.899 & 2007A & 0.4 & 10441(71) & 140(2) & NV & 0.946 \\ 
1999by & 1.8 & 8978(103) & 191(10) & 91bg & 1.899 & 2007S & -3.8 & 10625(118) & 168(9) & 91T & 1.096 \\ 
1999by & 2.8 & 8876(98) & 201(11) & 91bg & 1.899 & 2007S & -0.8 & 10211(115) & 145(8) & 91T & 1.096 \\ 
1999cc & -3.2 & 12177(182) & 186(10) & HV & 1.46 & 2007af & -3.7 & 10854(122) & 159(9) & NV & 1.2 \\ 
1999cc & -0.2 & 11604(169) & 182(10) & HV & 1.46 & 2007af & 0.3 & 10386(125) & 157(8) & NV & 1.2 \\ 
1999cl & -4.3 & 13081(138) & 217(11) & HV & 1.144 & 2007le & -10.4 & 14892(168) & 200(11) & HV & 1.015 \\ 
1999cl & 0.7 & 12135(128) & 207(11) & HV & 1.144 & 2007le & -8.4 & 14060(156) & 191(10) & HV & 1.015 \\ 
1999dq & -3.5 & 10485(111) & 140(8) & 91T & 0.961 & 2007le & -7.5 & 13649(147) & 185(10) & HV & 1.015 \\ 
1999dq & 0.5 & 10441(110) & 123(7) & 91T & 0.961 & 2007le & -6.5 & 13341(144) & 181(10) & HV & 1.015 \\ 
1999gp & -4.9 & 11918(149) & 204(11) & 91T & 0.908 & 2007le & -4.7 & 12764(135) & 170(9) & HV & 1.015 \\ 
1999gp & 0 & 10939(82) & 158(3) & 91T & 0.908 & 2007le & 2.3 & 12061(149) & 152(8) & HV & 1.015 \\ 
2000cx & -3.8 & 13856(79) & 163(4) & pec & 0.99 & 2007le & 3.3 & 12098(138) & 148(8) & HV & 1.015 \\ 
2000cx & 0.2 & 13692(53) & 138(2) & pec & 0.99 & 2007le & 4.3 & 12048(141) & 150(8) & HV & 1.015 \\ 
2000dk & -4.3 & 11606(131) & 179(10) & unclear & 1.718 & 2008Q & -3.3 & 11473(61) & 148(2) & NV & 1.25 \\ 
2000dk & 0.4 & 10675(118) & 172(9) & unclear & 1.718 & 2008Q & -0.3 & 11126(63) & 145(2) & NV & 1.25 \\ 
2001cp & -4 & 11218(137) & 146(8) & NV & 0.915 & 2008ar & -3.7 & 10497(157) & 149(8) & NV & 1.078 \\ 
2001cp & 0.9 & 10547(130) & 130(7) & NV & 0.915 & 2008ar & -0.7 & 10510(140) & 153(8) & NV & 1.078 \\ 
2001ep & -3.2 & 10671(118) & 168(9) & NV & 1.356 & 2011fe & -15 & 14401(48) & 228(2) & NV & 1.18 \\ 
2001ep & -0.2 & 10195(115) & 165(9) & NV & 1.356 & 2011fe & -14 & 13636(44) & 216(2) & NV & 1.18 \\ 
2002bo & -12.5 & 17000(196) & 221(12) & HV & 1.08 & 2011fe & -13 & 13097(45) & 201(2) & NV & 1.18 \\ 
2002bo & -8.4 & 15192(173) & 220(12) & HV & 1.08 & 2011fe & -12 & 12671(25) & 196(2) & NV & 1.18 \\ 
2002bo & -7.4 & 14857(160) & 221(12) & HV & 1.08 & 2011fe & -11 & 12262(39) & 186(2) & NV & 1.18 \\ 
2002bo & -6.3 & 14455(156) & 220(12) & HV & 1.08 & 2011fe & -10 & 11879(38) & 183(2) & NV & 1.18 \\ 
2002bo & -5.4 & 14172(154) & 216(11) & HV & 1.08 & 2011fe & -9 & 11526(50) & 175(2) & NV & 1.18 \\ 
2002bo & -4.4 & 13939(152) & 216(11) & HV & 1.08 & 2011fe & -8 & 11287(49) & 175(2) & NV & 1.18 \\ 
2002bo & -3.4 & 13681(151) & 213(11) & HV & 1.08 & 2011fe & -7 & 11067(55) & 165(2) & NV & 1.18 \\ 
2002bo & -2.5 & 13433(148) & 211(11) & HV & 1.08 & 2011fe & -6 & 10907(54) & 162(2) & NV & 1.18 \\ 
2002bo & -1.5 & 13217(145) & 208(11) & HV & 1.08 & 2011fe & -5 & 10815(47) & 160(2) & NV & 1.18 \\ 
2002cd & -3.7 & 15201(205) & 176(9) & unclear & 0.794 & 2011fe & -1 & 10386(64) & 150(2) & NV & 1.18 \\ 
2002cd & 0.3 & 14979(193) & 165(9) & unclear & 0.794 & 2011fe & 0 & 10296(85) & 148(3) & NV & 1.18 \\ 
2002er & -3.9 & 11740(131) & 159(8) & NV & 1.23 & 2011fe & 1 & 10220(146) & 147(8) & NV & 1.18 \\ 
2002er & 0.1 & 11491(129) & 167(9) & NV & 1.23 & 2011fe & 3 & 10089(156) & 145(8) & NV & 1.18 \\ 
2002he & -3.3 & 12641(141) & 169(9) & HV & 1.494 & 2011fe & 4 & 10066(167) & 147(8) & NV & 1.18 \\ 
2002he & 0.2 & 12155(134) & 167(9) & HV & 1.494 & 2011fe & -5 & 10815(47) & 159(2) & NV & 1.18 \\ 
2003W & -3.4 & 14445(236) & 226(7) & unclear & 1.113 & 2011fe & 0 & 10296(85) & 149(3) & NV & 1.18 \\ 
2003W & -0.3 & 14133(157) & 214(4) & unclear & 1.113 & 2012fr & -12.5 & 15500(160) & 283(2) & 99aa & 0.8 \\ 
2003cg & -4.1 & 11225(135) & 145(8) & NV & 1.22 & 2012fr & -11.5 & 14572(39) & 252(2) & 99aa & 0.8 \\ 
2003cg & -0.2 & 10851(127) & 141(8) & NV & 1.22 & 2012fr & -6.4 & 13131(55) & 203(2) & 99aa & 0.8 \\ 
2003du & -14.1 & 13210(71) & 200(11) & NV & 1.07 & 2012fr & -2.4 & 12905(147) & 183(7) & 99aa & 0.8 \\ 
2003du & -12.1 & 12103(89) & 195(4) & NV & 1.07 & 2012fr & -1.3 & 12969(165) & 181(7) & 99aa & 0.8 \\ 
2003du & -8.1 & 10973(45) & 162(2) & NV & 1.07 & 2012fr & -0.4 & 12936(169) & 178(7) & 99aa & 0.8 \\ 
2003du & -7.1 & 10632(56) & 157(2) & NV & 1.07 & 2012fr & 0 & 13195(175) & 167(6) & 99aa & 0.8 \\ 
2003du & -6.1 & 10773(141) & 163(9) & NV & 1.07 & 2012fr & 2 & 13338(194) & 156(6) & 99aa & 0.8 \\ 
2003du & -5.1 & 10413(81) & 147(2) & NV & 1.07 & 2012fr & 3 & 12996(137) & 150(3) & 99aa & 0.8 \\ 
2003du & -4.1 & 10474(131) & 155(8) & NV & 1.07 & 2014J & -11 & 14267(155) & 186(10) & HV & 1.022 \\ 
2003du & -3.1 & 10195(134) & 145(8) & NV & 1.07 & 2014J & -9 & 13619(150) & 183(10) & HV & 1.022 \\ 
2003du & -2.1 & 10280(117) & 143(8) & NV & 1.07 & 2014J & -7 & 12856(137) & 177(9) & HV & 1.022 \\ 
2003du & -1.1 & 10250(117) & 143(8) & NV & 1.07 & 2014J & -5 & 12642(135) & 167(9) & HV & 1.022 \\ 
2003du & -0.1 & 10133(119) & 137(7) & NV & 1.07 & 2014J & -3 & 12309(130) & 162(9) & HV & 1.022 \\ 
2003du & 0.9 & 10262(147) & 138(8) & NV & 1.07 & 2014J & -2 & 11910(126) & 160(9) & HV & 1.022 \\ 
2003du & 1.9 & 9789(116) & 138(7) & NV & 1.07 & 2014J & 0 & 12012(128) & 155(8) & HV & 1.022 \\ 
2003du & 2.9 & 10052(130) & 135(7) & NV & 1.07 & 2014J & 3 & 11930(129) & 147(8) & HV & 1.022 \\ 
2003du & 4.9 & 10115(129) & 132(7) & NV & 1.07 & 2014J & -5 & 12642(135) & 166(9) & HV & 1.022 \\ 
2003du & 5.9 & 10187(160) & 137(8) & NV & 1.07 & 2014J & 0 & 12012(128) & 155(8) & HV & 1.022 \\ 
2003du & -4.1 & 10474(131) & 155(8) & NV & 1.07 &  \\ 

	\hline
\multicolumn{12}{l}{Note: `$V^{Si6}$' and `$\gamma^{Si6}$' represent the velocity and FWHM of Si II $\lambda$6355, respectively. `$\Delta m_{15}$' represents the}\\
\multicolumn{12}{l}{B-band decline rate;}\\
\end{longtable}

\setlength{\tabcolsep}{6pt}
\begin{longtable}{ccccc|ccccc}
	\caption{The absorption depths and FWHMs of Si II $\lambda$5972 \label{Tab3}}\\
	\hline
	SN & Phase & $A^{Si5}$ & $\gamma^{Si5}$ & $\Delta m_{15}$ & SN & Phase & $A^{Si5}$ & $\gamma^{Si5}$ & $\Delta m_{15}$ \\
	~ &(days) & (km/s) &(\AA) & (mag) & ~ &(days) & (km/s) &(\AA) & (mag) \\
	\hline
	\endfirsthead
	
\multicolumn{4}{l}{\textbf{Table 3 continued}} \\
	\hline
SN & Phase & $A^{Si5}$ & $\gamma^{Si5}$ & $\Delta m_{15}$ & SN & Phase & $A^{Si5}$ & $\gamma^{Si5}$ & $\Delta m_{15}$ \\
~ &(days) & (km/s) &(\AA) & (mag) & ~ &(days) & (km/s) &(\AA) & (mag) \\
\hline
\endhead

\hline
\multicolumn{4}{l}{Continued on next page} \\
\endfoot	

\endlastfoot		
	
1981B & 0 & 0.1728(0.0046) & 181(10) & 0.91 & 2004as & -0.1 & 0.1447(0.0046) & 182(10) & 1.111 \\ 
1986G & 0 & 0.3655(0.0091) & 161(9) & 1.65 & 2004at & 0.2 & 0.1282(0.0042) & 88(6) & 1.091 \\ 
1989B & -1 & 0.1739(0.005) & 174(9) & 1.35 & 2004gs & -0.3 & 0.3601(0.0158) & 154(8) & 1.775 \\ 
1992A & 0 & 0.2353(0.0093) & 162(9) & 1.41 & 2005A & 0.6 & 0.0996(0.0029) & 161(9) & 1.222 \\ 
1994ae & 0 & 0.123(0.0029) & 137(7) & 0.96 & 2005M & -0.1 & 0.1396(0.0037) & 168(9) & 0.827 \\ 
1995E & -0.2 & 0.2178(0.007) & 160(9) & 1.16 & 2005cf & 0.3 & 0.1498(0.0039) & 86(7) & 1.1 \\ 
1996X & -0.1 & 0.1315(0.0056) & 138(6) & 1.26 & 2005eq & 0.4 & 0.0989(0.0024) & 128(7) & 0.882 \\ 
1996ai & 0.1 & 0.098(0.0033) & 134(7) & 1 & 2005ke & -0.3 & 0.3942(0.0099) & 158(8) & 1.762 \\ 
1997bp & -0.2 & 0.0797(0.0019) & 119(6) & 1.08 & 2005na & 0.5 & 0.1094(0.0039) & 136(7) & 1.239 \\ 
1997dt & 0.2 & 0.0957(0.0023) & 106(6) & 1.04 & 2006N & 0.7 & 0.1608(0.0042) & 138(7) & 1.57 \\ 
1998aq & -0.2 & 0.1075(0.0028) & 126(7) & 1.11 & 2006S & 0.3 & 0.0512(0.0015) & 135(7) & 0.9 \\ 
1998bp & -2.9 & 0.4051(0.012) & 118(6) & 1.83 & 2006ax & 0.2 & 0.1268(0.0022) & 143(7) & 1.016 \\ 
1998bp & -1.9 & 0.4027(0.0106) & 120(7) & 1.83 & 2006bt & -0.1 & 0.3367(0.0127) & 176(9) & 0.877 \\ 
1998bp & -0.9 & 0.4091(0.0107) & 122(7) & 1.83 & 2006cj & -0.2 & 0.1785(0.0207) & 142(9) & 0.81 \\ 
1998bp & 0.1 & 0.407(0.0099) & 121(7) & 1.83 & 2006gt & -0.1 & 0.4231(0.029) & 121(3) & 1.66 \\ 
1998bp & 1 & 0.4015(0.01) & 125(7) & 1.83 & 2006sr & -1 & 0.1795(0.0051) & 158(8) & 1.39 \\ 
1998de & -0.3 & 0.4557(0.0184) & 152(8) & 1.93 & 2007A & 0.4 & 0.1224(0.0038) & 128(6) & 0.946 \\ 
1998dh & -0.5 & 0.1947(0.0048) & 130(7) & 1.227 & 2007F & -0.9 & 0.1101(0.0042) & 136(7) & 1.03 \\ 
1998dx & -0.7 & 0.1943(0.0164) & 133(5) & 1.55 & 2007S & -0.8 & 0.0664(0.0017) & 145(8) & 1.096 \\ 
1998es & -0.5 & 0.0803(0.0019) & 137(7) & 0.978 & 2007af & 0.3 & 0.1989(0.0059) & 129(7) & 1.2 \\ 
1999aa & -11.4 & 0.1216(0.0032) & 233(11) & 0.94 & 2007bc & 0.4 & 0.3042(0.0084) & 153(8) & 1.349 \\ 
1999aa & -10 & 0.1232(0.0022) & 286(15) & 0.94 & 2007ci & -8.2 & 0.3839(0.0175) & 174(5) & 1.744 \\ 
1999aa & -9.1 & 0.1274(0.0019) & 257(13) & 0.94 & 2007ci & -7.2 & 0.423(0.0167) & 182(10) & 1.744 \\ 
1999aa & -8.1 & 0.1197(0.002) & 238(12) & 0.94 & 2007ci & -6.2 & 0.3698(0.0133) & 182(10) & 1.744 \\ 
1999aa & -7.2 & 0.1127(0.0022) & 226(12) & 0.94 & 2007ci & -5.2 & 0.3574(0.0101) & 170(9) & 1.744 \\ 
1999aa & -6.1 & 0.1132(0.0023) & 199(11) & 0.94 & 2007ci & -4.2 & 0.324(0.0087) & 158(8) & 1.744 \\ 
1999aa & -5.1 & 0.0902(0.0022) & 168(9) & 0.94 & 2007ci & -2.2 & 0.3165(0.0092) & 162(9) & 1.744 \\ 
1999aa & -4.1 & 0.0766(0.0017) & 150(8) & 0.94 & 2007ci & -1.2 & 0.3123(0.0098) & 158(9) & 1.744 \\ 
1999aa & -3.1 & 0.0754(0.0017) & 145(8) & 0.94 & 2007co & -0.2 & 0.1326(0.0049) & 135(7) & 1.04 \\ 
1999aa & -2.1 & 0.0727(0.0017) & 137(7) & 0.94 & 2007hj & -2.6 & 0.4286(0.0115) & 152(8) & 1.953 \\ 
1999aa & -0.1 & 0.0837(0.002) & 132(7) & 0.94 & 2007hj & -1.5 & 0.4073(0.0104) & 145(8) & 1.953 \\ 
1999ac & -1 & 0.1252(0.0023) & 161(4) & 1.182 & 2007hj & -0.6 & 0.4193(0.011) & 146(8) & 1.953 \\ 
1999cc & -0.2 & 0.2388(0.0096) & 143(8) & 1.46 & 2007hj & 1.4 & 0.3745(0.0103) & 141(8) & 1.953 \\ 
1999cl & 0.7 & 0.1609(0.0037) & 171(9) & 1.144 & 2007hj & 2.4 & 0.4017(0.0114) & 141(8) & 1.953 \\ 
1999dq & 0.5 & 0.0832(0.0019) & 173(9) & 0.961 & 2007hj & 3.5 & 0.3799(0.0091) & 144(8) & 1.953 \\ 
1999ee & -0.4 & 0.1068(0.0025) & 189(10) & 0.96 & 2007hj & 4.4 & 0.3766(0.0094) & 137(7) & 1.953 \\ 
1999gp & 0 & 0.0324(0.0017) & 143(11) & 0.908 & 2008Q & -0.3 & 0.1777(0.0087) & 90(9) & 1.25 \\ 
2000cx & 0.2 & 0.1073(0.0027) & 138(7) & 0.99 & 2008ar & -0.7 & 0.1425(0.005) & 90(5) & 1.078 \\ 
2000dk & 0.4 & 0.3554(0.0089) & 135(7) & 1.718 & 2008bf & 0.7 & 0.0917(0.0027) & 116(6) & 1.029 \\ 
2001cp & 0.9 & 0.0719(0.0023) & 91(5) & 0.915 & 2011fe & -15 & 0.2418(0.0032) & 192(4) & 1.18 \\ 
2001da & 0 & 0.19(0.0052) & 156(8) & 1.23 & 2011fe & -14 & 0.2937(0.0035) & 187(6) & 1.18 \\ 
2001eh & -0.1 & 0.0573(0.0036) & 131(17) & 0.811 & 2011fe & -13 & 0.2841(0.0034) & 179(4) & 1.18 \\ 
2001ep & -0.2 & 0.2792(0.0073) & 165(9) & 1.356 & 2011fe & -12 & 0.2929(0.0023) & 185(4) & 1.18 \\ 
2001fe & 0.3 & 0.0947(0.0022) & 123(7) & 1.03 & 2011fe & -11 & 0.2792(0.0031) & 181(4) & 1.18 \\ 
2002cd & 0.3 & 0.055(0.0018) & 105(6) & 0.794 & 2011fe & -10 & 0.266(0.003) & 182(4) & 1.18 \\ 
2002cr & 0.7 & 0.2311(0.0067) & 129(7) & 1.229 & 2011fe & -9 & 0.2447(0.0037) & 175(9) & 1.18 \\ 
2002er & 0.1 & 0.1492(0.0039) & 123(7) & 1.23 & 2011fe & -8 & 0.2386(0.0058) & 150(8) & 1.18 \\ 
2002he & -8.3 & 0.2962(0.017) & 163(5) & 1.494 & 2011fe & -7 & 0.216(0.0037) & 165(9) & 1.18 \\ 
2002he & -6.2 & 0.2317(0.0082) & 148(8) & 1.494 & 2011fe & -6 & 0.2104(0.0036) & 161(9) & 1.18 \\ 
2002he & -3.3 & 0.1927(0.0049) & 138(7) & 1.494 & 2011fe & -5 & 0.2124(0.0032) & 159(8) & 1.18 \\ 
2002he & -1.2 & 0.1668(0.0048) & 122(7) & 1.494 & 2011fe & -1 & 0.176(0.0041) & 135(6) & 1.18 \\ 
2002he & 0.2 & 0.1921(0.0048) & 140(8) & 1.494 & 2011fe & 0 & 0.1768(0.0053) & 134(7) & 1.18 \\ 
2002he & 3.2 & 0.2214(0.0054) & 159(8) & 1.494 & 2011fe & 1 & 0.1736(0.0068) & 131(7) & 1.18 \\ 
2002he & 5.7 & 0.2155(0.0104) & 108(7) & 1.494 & 2011fe & 3 & 0.1674(0.0073) & 143(8) & 1.18 \\ 
2002kf & 0.8 & 0.1924(0.0063) & 133(7) & 1.2 & 2011fe & 4 & 0.1916(0.0093) & 147(8) & 1.18 \\ 
2003W & -0.3 & 0.0607(0.002) & 174(7) & 1.113 & 2012fr & -12.5 & 0.0908(0.002) & 192(9) & 0.8 \\ 
2003cg & -0.2 & 0.1701(0.0048) & 102(6) & 1.22 & 2012fr & -11.5 & 0.0801(0.0012) & 178(9) & 0.8 \\ 
2003ch & -0.2 & 0.1703(0.0106) & 120(4) & 1.22 & 2012fr & -6.4 & 0.0797(0.002) & 103(4) & 0.8 \\ 
2003cq & 0.1 & 0.2289(0.0294) & 162(8) & 1.26 & 2012fr & -2.4 & 0.0664(0.0018) & 110(6) & 0.8 \\ 
2003du & -0.1 & 0.0965(0.0027) & 109(6) & 1.07 & 2012fr & -1.3 & 0.0647(0.0021) & 120(7) & 0.8 \\ 
2003it & -0.2 & 0.1589(0.0039) & 123(4) & 1.36 & 2012fr & -0.4 & 0.0777(0.0027) & 126(7) & 0.8 \\ 
2003iv & 0.1 & 0.253(0.0135) & 121(4) & 1.65 & 2014J & 0 & 0.1131(0.0026) & 127(7) & 1.022 \\ 
	\hline
\multicolumn{10}{l}{Note: `$A^{Si5}$' represents the absorption depth of Si II $\lambda$5972, `$\gamma^{Si6}$' represents the FWHM of Si II $\lambda$5972.}\\
\end{longtable}


\begin{thebibliography}{99}	
	\bibitem[Benetti et al. (2005)]{Benetti05} Benetti S. et al., 2005, \apj, 623,1011

	\bibitem[Blondin et al. (2012)]{Blondin12} Blondin S. et al., 2012, \aj, 143, 126

	\bibitem[Blondin et al. (2018)]{Blondin18} Blondin S., Dessart L., Hillier D. J., 2018, \mnras, 474, 3931

    \bibitem[Branch et al. (2006)]{Branch06} Branch, D., et al. 2006, PASP, 118, 560

	\bibitem[Branch et al. (2009)]{Branch09} Branch D., Dang L. C., Baron E., 2009, PASP, 121, 238

	\bibitem[Childress et al. (2013)]{Childress13} Childress, M. J., Scalzo, R. A., Sim, S., et al. 2013, \apj, 770, 29

	\bibitem[Childress et al. (2014)]{Childress14} Childress M. J., Filippenko A. V., Ganeshalingam M., Schmidt B. P., 2014, \mnras, 437, 338

	\bibitem[Filippenko et al. (1992a)]{Filippenko92a}Filippenko A. V. et al., 1992, Astron. J., 104, 1543

	\bibitem[Filippenko et al. (1992b)]{Filippenko92b}Filippenko A. V. et al., 1992, Astrophys. J., 384, L15

    \bibitem[Fl\"{o}rs et al. (2020)]{Flors20} Fl\"{o}rs A. et al., 2020, \mnras, 491, 2902 

	\bibitem[Folatelli et al. (2013)]{Folatelli13} Folatelli G. et al., 2013, \apj, 773, 53

	\bibitem[Freedman et al. (2019)]{Freedman19} Freedman W. L. et al., 2019, \apj, 882, 34

	\bibitem[Hakobyan et al. (2025)]{Hakobyan25} Hakobyan A. A., Gevorgyan M. H., Karapetyan A. G. et. al., 2025, arXiv:2511.23305

	\bibitem[Hillebrandt \& Niemeyer (2000)]{Hillebrandt00} Hillebrandt W., Niemeyer J. C., 2000, \araa, 38, 191

	\bibitem[Iben \& Tutukov (1984)]{Iben84} Iben I., Tutukov A. V., 1984, \apjs, 54, 355

	\bibitem[Jha et al. (2019)]{Jha19} Jha S. W., Maguire K., Sullivan M., 2019, Nature Astron., 3, 706

	\bibitem[Ruiz-Lapuente et al. (2023)]{Ruiz-Lapuente23} Ruiz-Lapuente P., Gonz\'{a}lez Hern\'{a}ndez, J. I., Cartieret R. et. al., 2023, \apj 947 90

	\bibitem[Khokhlov (1991)]{Khokhlov91} Khokhlov, A. M., 1991, \aap, 245, 114

	\bibitem[Kushnir et al. (2013)]{Kushnir13} Kushnir D., Katz B., Dong S., Livne E., Fern\'{a}ndez R., 2013, \apj, 778, L37

	\bibitem[Li et al. (2001)]{Li01} Li W., Filippenko A. V., Treffers R. R., Riess A. G., Hu J., \& Qiu Y., 2001, ApJ, 546, 734

	\bibitem[Li et al. (2011)]{Li11} Li W. et al., 2011, \nat, 480, 348

	\bibitem[Maeda et al. (2010)]{Maeda10} Maeda K. et al., 2010, \nat, 466, 82

	\bibitem[Maeda et al. (2018)]{Maeda18} Maeda K., Jiang J., Shigeyama T., Doi M., 2018, \apj, 861, 78

	\bibitem[Maguire et al. (2014)]{Maguire14} Maguire K. et al., 2014, \mnras, 444, 3258

	\bibitem[Matheson et al. (2008)]{Matheson08} Matheson T. et al., 2008, \aj, 135, 1598

	\bibitem[Ni et al. (2023)]{Ni23} Ni Y. et al, 2023, \apj, 959, 132

	\bibitem[Nomoto (1982)]{Nomoto82} Nomoto K., 1982, \apj, 253, 798

	\bibitem[Nugent et al. (1995)]{Nugent95} Nugent P., Phillips M., Baron E., Branch D., Hauschildt P. 1995, \apjl, 455, L147

	\bibitem[Pakmor et al. (2013)]{Pakmor13} Pakmor R., Kromer M., Taubenberger S., Springel V., 2013, \apj, 770, L8

	\bibitem[Perlmutter et al. (1999)]{Perlmutter99} Perlmutter S. et al., 1999, \apj, 517, 565

	\bibitem[Phillips (1992)]{Phillips92}Phillips M. M. et al.,1992, Astron. J., 103, 1632	

	\bibitem[Phillips (1993)]{Phillips93} Phillips M. M., 1993, \apjl, 413, L105

	\bibitem[Phillips et al. (1999)]{Phillips99} Phillips M. M. et al., 1999, \aj, 118, 1766

	\bibitem[Riess et al. (1998)]{Riess98} Riess A. G. et al., 1998, \aj, 116, 1009

	\bibitem[Riess et al. (2019)]{Riess19} Riess A. G., Casertano S., YuanW., Macri L. M. Scolnic D., 2019, \apj, 876, 85

	\bibitem[Shen et al. (2018)]{Shen18} Shen K. J., Kasen D., Miles B. J., Townsley D. M., 2018, \apj, 854, 52

	\bibitem[Silverman et al. (2012a)]{Silverman12a} Silverman J. M. et al., 2012a, \mnras, 425, 1789

	\bibitem[Silverman et al. (2012b)]{Silverman12b} Silverman J. M., Kong J. J., Filippenko A. V., 2012b, \mnras, 425, 1819

	\bibitem[Silverman et al. (2015)]{Silverman15} Silverman J. M. et al., 2015, \mnras, 451, 1973

	\bibitem[Wang et al. (2009)]{Wang09} Wang X. et al., 2009, \apjl, 699, L139

	\bibitem[Webbink (1984)]{Webbink84} Webbink R. F., 1984, \apj, 277, 355

	\bibitem[Whelan \& Iben (1973)]{Whelan73} Whelan J., Iben I., 1973, \apj, 186,1007

	\bibitem[Meng et al. (2019)]{Meng19} Meng X.,2019, \apj, 886, 58

	\bibitem[Zhang et al. (2014)]{Zhangjj14} Zhang J., Wang X., Bai J. et al., 2014, \aj, 148, 1
	
	\bibitem[Zhao et al. (2015)]{Zhao15} Zhao X. et al., 2015, \apjs, 220, 20

	\bibitem[Zhao et al. (2016)]{Zhao16} Zhao X. et al., 2016, \apj, 826, 211

	\bibitem[Zhao et al. (2021)]{Zhao21} Zhao X. et al., 2021, \mnras, 503, 4667

	\bibitem[Zhao et al. (2024)]{Zhao24} Zhao X. et al., 2024, \mnras, 535, 3470
	
\end{thebibliography}
\end{document}